# Hybrid ferroelectric tunnel junctions:
# State-of-the-art, challenges and opportunities

King-Fa Luo[1], Zhijun Ma[2*], Daniel Sando[1, 3], Qi Zhang[1,4*], Nagarajan Valanoor[1*]

[1]*School of Materials Science and Engineering, University of New South Wales, Sydney NSW 2052*
[2]*Ministry of Education Key Laboratory for the Synthesis and Application of Organic Functional Molecules, Hubei Key Laboratory for Precision Synthesis of Small Molecule Pharmaceuticals, Hubei Key Laboratory of Micro-Nanoelectronic Materials and Devices, Hubei University, Wuhan 430062, China*
[3]*School of Physical and Chemical Sciences, University of Canterbury, Christchurch 8140, New Zealand*
[4]*CSIRO, Manufacturing, Lindfield, NSW 2070, Australia*

Correspondence: mazhijun@hubu.edu.cn; peggy.zhang@csiro.au; nagarajan@unsw.edu.au

## Abstract

Ferroelectric tunnel junctions (FTJs) harness the unique combination of ferroelectricity and quantum tunneling, and thus herald new opportunities in next-generation nonvolatile memory technologies. Recent advancements in the fabrication of ultrathin heterostructures have enabled the integration of ferroelectrics with various functional materials, forming hybrid tunneling-diode junctions. These junctions benefit from the modulation of the functional layer/ferroelectric interface through ferroelectric polarization, thus enabling further modalities and functional capabilities than in addition to tunneling electroresistance. This perspective aims to provide in-depth insight into novel physical phenomena of several typical ferroelectric hybrid junctions, ranging from ferroelectric/dielectric, ferroelectric/multiferroic, ferroelectric/superconducting to ferroelectric/2D materials, and finally their expansion into the realm of ferroelectric resonant tunneling diodes (FeRTDs). This latter aspect, *i.e.,* resonant tunneling offers a radically new approach to exploiting tunneling behavior in ferroelectric heterostructures. We discuss examples that have successfully shown room temperature ferroelectric control of parameters such as the resonant peak, tunnel current ratio at peak and negative differential resistance. We conclude the perspective by summarizing the challenges and highlighting the opportunities for the future development of hybrid FTJs with a special emphasis on a new possible type of FeRTD device. The prospects for enhanced performance and expanded functionality ignite tremendous excitement in hybrid FTJs and FeRTDs for future nanoelectronics.

**Introduction**

The so-called tunnel junction refers to an ultrathin insulating layer sandwiched between two metals. According to quantum mechanics, electrons can "tunnel" through the insulator with a certain probability, even if the energy of the electrons is less than the potential barrier imposed by the insulator.[1–3] Originally, with such tunnel junctions, their added functionality arises from the metal electrodes, whether it be magnetic or superconducting tunnel junctions.[4,5] The sandwiched layer itself is a non-polar dielectric, typically unperturbed by external stimuli. This approach changed with the introduction of a ferroelectric layer as the tunnel barrier.[6]

Ferroelectricity is a property of certain crystalline materials that possess a spontaneous electric polarization which can be reversed by an applied external electric field.[7] In the context of data storage technologies, ferroelectrics have been widely used in devices such as ferroelectric random access memories (FeRAMs) and ferroelectric field-effect transistors (FeFETs).[9,10] With the advances in the processing of complex ferroelectric oxides and silicon integration techniques in the 1980s, a paradigm shift in the science and engineering of ferroelectrics away from the dominance of single crystal and bulk ferroelectric devices into the realm of thin films was evoked.[11] The further requirement for size reductions imposed by the semiconductor industry gave rise to the strong interest concerning the possible existence of a critical thickness for ferroelectricity.[12,13] Encouragingly, theoretical and experimental efforts of the last two decades have shown that ferroelectricity can exist in thin films with the thickness down to just a few unit cells.[14–17] These findings opened up a tantalizing new opportunity - could one exploit the combination of ferroelectricity with the quantum-mechanical tunneling effect to realize new types of memory effects?

When the ferroelectric is sandwiched to form a (asymmetric) tunnel junction, an applied electric field not only switches its polarization, it also effectively affects the averaged barrier height of the tunnel junction. As a result it gives rise to a difference in the transmission of electrons for different polarization orientations which in turn imparts different resistance values to the junction.[8] This effect sketched in Figure 1 is known as the tunneling electroresistance (TER) effect.[8,18]

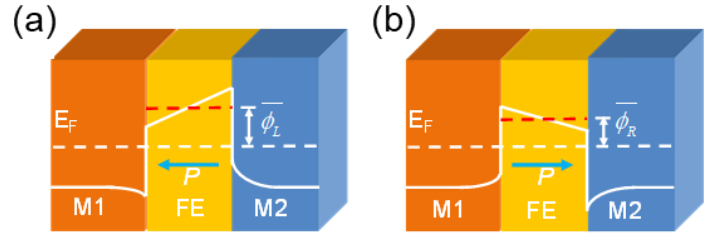

**Figure 1.** Schematics of the structures and potential energy profiles of a typical Metal (M1)/Ferroelectric (FE)/Metal (M2) FTJ. a,b) show the ferroelectric polarization pointing to the left and right, respectively. $\overline{\phi}_L$ and $\overline{\phi}_R$ indicate the difference in average barrier height, and $E_F$ is the Fermi level.

In TER, one exploits the interplay between electron tunneling and ferroelectric polarization. This then leads to the notion of polarization control of current in ferroelectric-based quantum-tunneling devices.[6,19,20] Polarization control of tunneling involves the following key aspects:

(1) **Ferroelectric polarization amplitude**. Robust ferroelectricity can amplify the asymmetry of the potential energy profile, leading to a substantial TER. In general, a compressive strain imposed by the substrate is desirable for ABO$_3$-type perovskite ferroelectric films that possess polarization in the c-direction.[21–23] For example, epitaxial thin film BiFeO$_3$ (BFO) grown on substrates that induce a large compressive strain can be stabilized as the "T-like phase" with a giant axial ratio of c/a=1.2.[24] This so-called "T-phase" BFO possesses a large polarization on account of its increased tetragonality, resulting in a large OFF/ON ratio >10, 000 as shown in Co/BFO/Ca$_{0.96}$Ce$_{0.04}$MnO$_3$ FTJs on (001) YAlO$_3$ substrates.[24]

(2) **Ferroelectric field effect-induced band bending at ferroelectric/semiconductor interface.** The major carriers in a semiconductor electrode accumulate or deplete at the ferroelectric/semiconductor interface, depending on the orientation of the ferroelectric polarization. As the majority carriers – electrons – are depleted, the band at the ferroelectric/semiconductor interface bends upwards and both the effective barrier width and height increase, thereby significantly affecting electron tunneling and the resultant device resistance.[25–27]

(3) **Memristive behavior.** The ferroelectric polarization can be continuously modulated by externally applied voltages, explained in terms of the evolution of domain structure.[28] The response of ferroelectric polarization to applied voltages enables the realization of ferroelectric tunnel memristors, which are promising candidates for high-density storage and neuromorphic networks.[29,30]

(4) **Metal-insulator phase transition**. The ferroelectric polarization can trigger the metal-insulator phase

transition in correlated electron oxides and redox reaction in superconducting junctions, both providing new mechanisms associated with the TER effect.[31–33]

Several excellent review articles have been published over the years on the fundamental physics and mechanisms of ferroelectric tunneling and "simple" ferroelectric tunnel junctions.[34,35] In this review, we put the spotlight on the next advanced concept, *viz*. ferroelectric devices that are based on hybrid tunneling concepts (explained next). The paper is organized as follows. Section 1 introduces the definition of hybrid ferroelectric junction. In Section 2, we explore various types of hybrid ferroelectric tunneling junctions (FTJs), including ferroelectric/dielectric, ferroelectric/multiferroic, ferroelectric/superconducting, and 2D/ferroelectric hybrid junctions. This is followed by a thorough review of ferroelectric resonant tunneling diodes (FeRTDs) in Section 3. We then present our perspective and outlook in Section 4, and summarize the review in Section 5.

1. What do we mean by a "hybrid" ferroelectric tunneling junction?

We define hybrid ferroelectric tunneling as electrons tunneling through a functional layer in series with an ultrathin ferroelectric film, as shown in Figure 2. The functional layer can be a paraelectric, dielectric insulator, ferromagnet, quantum well or interfacial layer between the ferroelectric and a semiconductor, and so on. The polarization control of hybrid tunneling not only provides means to tune the performance, but also opens many other possibilities for exploiting novel physical phenomena and new applications in future oxide electronics, as detailed next.

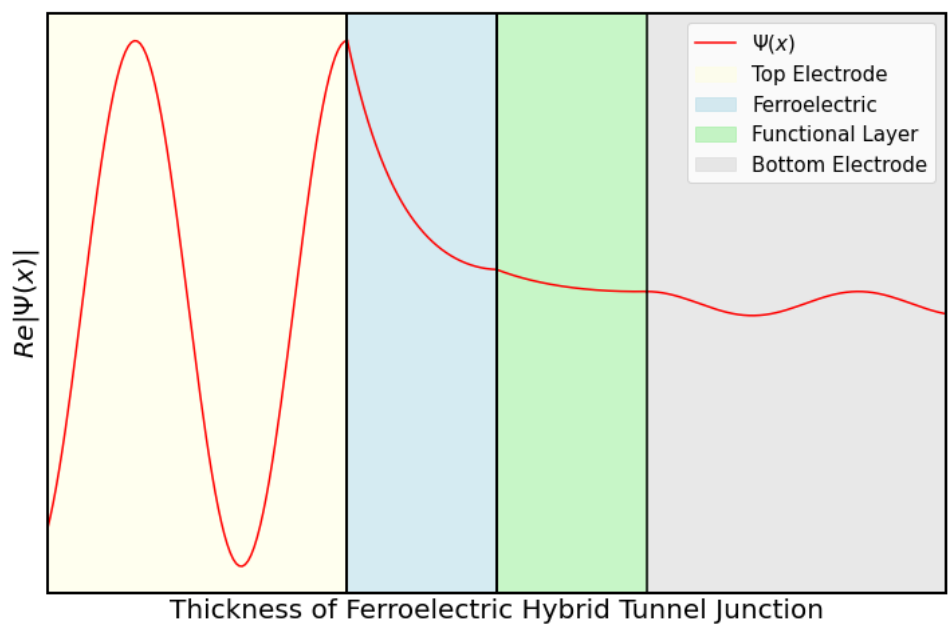

**Figure 2.** Schematic illustration of electron wavefunction tunneling across a prototypical hybrid ferroelectric tunnel junction.

## 2.1 Ferroelectric/dielectric hybrid junctions: Some typical varieties and examples

### a. Perovskite oxides-based hybrid junctions

As stated above, asymmetry in an FTJ is a key requirement to obtain the TER effect. FTJs with a single barrier and identical electrodes should theoretically have no TER effect, as the symmetric potential profile leads to no difference in average barrier height for the two opposite polarization states. To address this constraint, Tsymbal *et al.* proposed an FTJ with a layered composite barrier that consisted of a functional ferroelectric film and a thin film of a non-polar dielectric material (Figure 3a).[36] This non-polar oxide layer served as a switch that changed the barrier height of the FTJ from a low to high value when the polarization of the ferroelectric barrier is reversed. For example, in a MgO/BaTiO$_3$ (BTO) composite barrier, the non-polar MgO had a large amplitude of the tunneling barrier height, the conductance changed steeply as a function of MgO thickness, resulting in giant TER values.[36]

The performance of FTJs with a single barrier could be compromised by barrier thickness limits.[19,31,37] With an ultrathin barrier, the film leakage current crucially impacts the tunneling behavior. On the other hand, in the case of a thicker barrier, the tunneling current may be too small to be detected experimentally. For example, in an FTJ with BTO/SrTiO$_3$ (STO) composite barriers,[38] the extra STO barrier provided an additional degree of freedom to modulate the barrier potential and tunneling current (Figure 3b). This effectively overcame the fundamental barrier thickness limits of FTJs with a single barrier. In addition, as the total thickness of the composite barrier was fixed to 10 unit cells, the best performance was obtained for the barrier composition of BTO (6 unit cells)/STO (4 unit cells). The corresponding TER was more than 10 times higher than that of the single-BTO-barrier-FTJ with the same total barrier thickness. Ma *et al.* had also fabricated FTJs with composite BTO (2 nm)/STO (1~4 nm) barriers,[39] where they found that the largest TER effect was observed for the 3 nm-thick STO. This was attributed to the *weak* Fowler–Nordheim tunneling-assisted low OFF-state current relative to the 4 nm-thick counterparts.

### b. Two-dimensional electron gas (2DEG)-based hybrid junctions

Recent studies on a specific ferroelectric/dielectric junction have uncovered the formation of a two-dimensional electron gas (2DEG) or two-dimensional hole gas (2DHG) at certain oxide interfaces.[40–45] It was suggested that the formation of a 2DEG (2DHG) at the oxide interface could make FTJs with a ferroelectric/dielectric composite barrier a favorable candidate for nonvolatile memory applications.

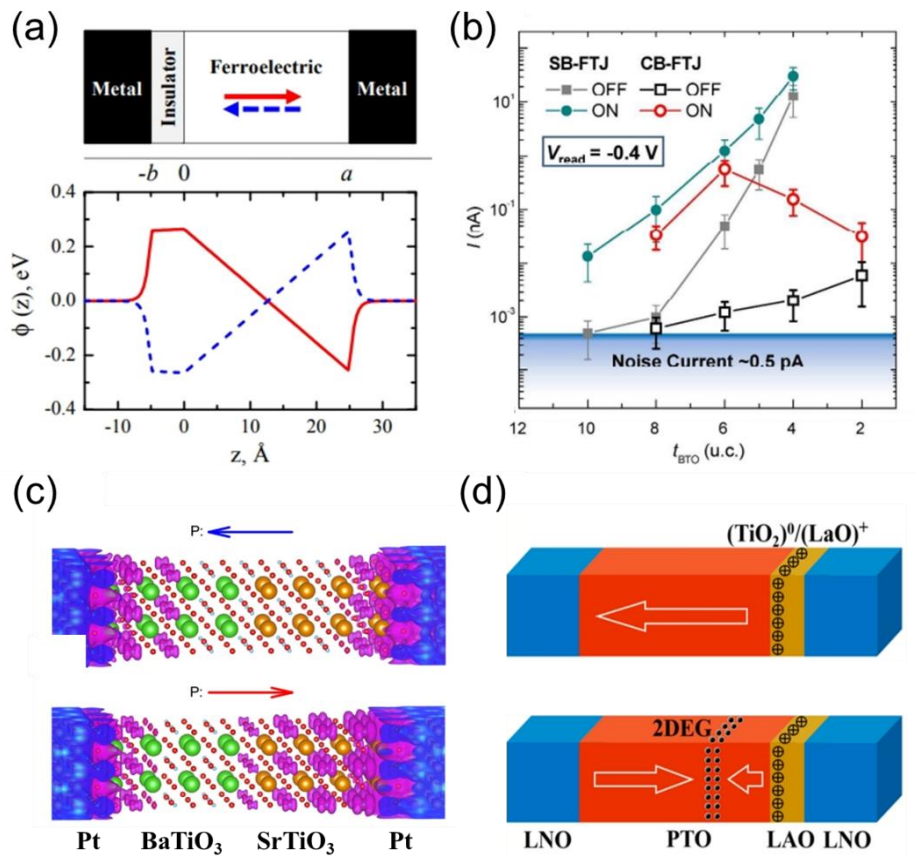

**Figure 3.** a) Geometry and the electrostatic potential profile of an FTJ with a composite barrier for two opposite polarization orientations. Reproduced with permission.[36] Copyright 2009, American Institute of Physics. b) BTO thickness-dependent ON- an OFF-state currents of Ti/BTO/SrRuO$_3$ (SRO) and Ti/BTO/STO/SRO FTJs. Reproduced with permission.[38] Copyright 2016, American Chemical Society. c) Charge density at the Fermi level in a Pt/BTO/STO/Pt FTJ indicating no 2DEG formed for the left polarization while a 2DEG formed at the interface and STO part for the right polarization. Reproduced with permission[46]. Copyright 2016, American Physics Society. d) Sketch of the (TiO$_2$)$^0$/(LaO)$^+$ terminated LaNiO$_3$/PbTiO$_3$/LaAlO$_3$/LaNiO$_3$ tunnel junction for the P← (top panel) and P→ (bottom panel) states. Symbols denote the positively charged interfacial (LaO)$^+$ atomic layer. The black small circles indicate the 2DEG formed at the head-to-head domain wall. Reproduced with permission.[47] Copyright 2019, American Chemical Society.

First-principles calculations by Wu *et al.* showed that the ferroelectric polarization induced a switchable 2DEG at the non-polar STO/BTO interface in Pt/BTO/STO/Pt FTJs.[46] When the ferroelectric polarization pointed from BTO to STO, a 2DEG formed at the interface and acted as a bridge for electrons to tunnel through the junctions, whereas no 2DEG existed at the interface after polarization reversal (Figure 3c). By utilizing the ferroelectrically switchable 2DEG, Pt/BTO/STO/Pt FTJs exhibited not only a giant TER but also a low resistance area (RA) product, which makes them suitable for practical applications. Though FTJs with a composite BTO/STO barrier showed a very large OFF/ON resistance ratio of $10^3 \sim 10^4$ orders in experimental studies, the ON state RA product was too high (50~200 MΩ μm$^2$ for a composite barrier thickness of ~4 nm),[38] suggesting that there was likely no 2DEG formed at the BTO/STO interface. In addition, Yang et al. predicted a TER effect in FTJs with a composite ferroelectric (PbTiO$_3$)/polar dielectric (LaAlO$_3$) layer.[47] Due to the formation of a 2DEG or 2DHG at the PbTiO$_3$/LaAlO$_3$ interface, the FTJs exhibited an OFF/ON resistance ratio exceeding a factor of $10^4$ and ON state RA product as low as about 1 kΩ μm$^2$. This 2DEG (2DHG) depended on the termination of the LaAlO$_3$ layer being (LaO)$^+$ or (AlO$_2$)$^-$ respectively, and could be switched ON and OFF by tuning the ferroelectric polarization of PbTiO$_3$ (Figure 3d).

c. Hafnia-based hybrid junctions

HfO$_2$-based ferroelectric films have attracted significant attention since they are complementary metal-oxide-semiconductor (CMOS) process compatible and lead-free. The discovery of impressive ferroelectric properties in doped fluorite-structure HfO$_2$ thin films in 2011 triggered an entirely new research theme.[48–52] The abnormal size effect, viz. enhanced ferroelectricity with scaling down the film thickness, makes HfO$_2$-based ferroelectrics an ideal barrier in FTJs. Single-barrier FTJs on Si or Ge substate have exhibited a relative low TER or OFF/ON resistance (current) ratio typically no more than 100.[53–56] FTJs combining a HfO$_2$-based ferroelectric film with a dielectric layer have thus been explored to enhance the performance. These dielectric layers include SiO$_2$,[57,58] TiO$_2$,[59] ZrO$_2$,[60,61] Al$_2$O$_3$,[62–64] among others.[65–68] Cheema et al. fabricated FTJs with a composite Zr:HfO$_2$(HZO)/SiO$_2$ layer by atomic layer deposition (ALD) directly on silicon substrates (Figure 4a).[69] The polar orthorhombic structure ($Pca2_1$) of the 1 nm HZO layer was confirmed by synchrotron X-ray characterization. Owing to the ultrathin barrier thickness, the HZO/SiO$_2$ based FTJs demonstrated large tunnel current density (> 1 A cm$^{-2}$ at low read voltage) and large polarization-driven TER (exceeding 200), orders of magnitude larger than reported thicker HfO$_2$-based FTJs.[70] In 2021, TER values in orders of ~$10^6$ were achieved theoretically in FTJs with a ferroelectric HfO$_2$/SiO$_2$ composite barrier by Chang et al., using a calibrated model in which electrons tunneling from conduction band (CBE) and valence band (VBE), and hole tunneling from the valence band (VBH) were considered.[71] In the following year, a TER of ~$10^6$ was realized experimentally in FTJs where Y-doped HfO$_2$ (YHO) was directly integrated on a heavily doped n-type (001)-Si substrate (Figure 4b).[72] The ultrathin YHO (~1 nm) exhibited a clear epitaxial relation with the Si substrate even in the presence of a thermally grown SiO$_x$ layer (~1 nm). The robust ferroelectricity (0.25 C m$^{-2}$) in epitaxial YHO, n-type semiconductor nature of the Si substrate, along with the enhanced asymmetry of the potential energy profile with the YHO/SiO$_x$ composite layer were collectively attributed as causes for the colossal TER.

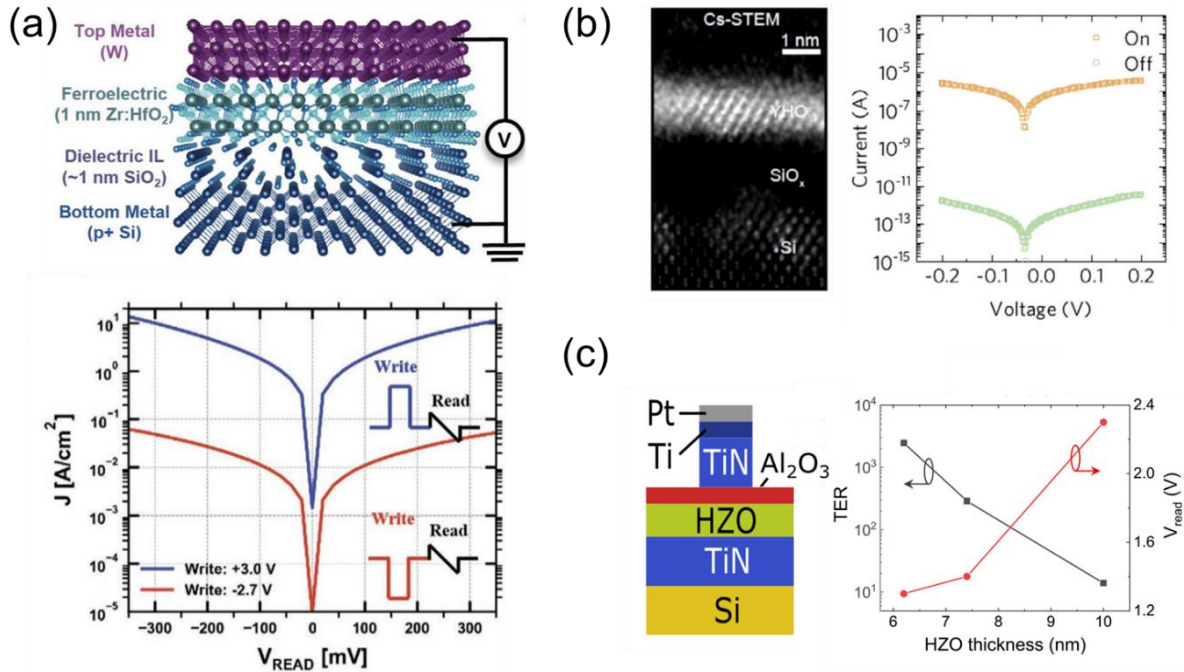

**Figure 4.** a) Schematic of a W/Zr:HfO$_2$/SiO$_2$/Si FTJ (top panel) and I-V measurements (bottom panel) after the indicated write voltage (+3.0 V, −2.7 V) was applied to set the Zr:HfO$_2$ polarization (FTJ current) into its respective state. Reproduced with permission.[69] Copyright 2021, Wiley-VCH GmbH. b) High-resolution scanning transmission electron microscopy cross-sectional image of the YHO/SiO$_x$/Si epitaxial structure (left panel) and current-voltage plot of On/Off states (right panel) of a TiN/YHO/SiOx/Si FTJ. Reproduced with permission.[72] Copyright 2021, Elsevier Ltd. c) Schematic of a TiN/HZO/Al$_2$O$_3$/Si FTJ (left panel) and TER and the optimal read voltage as a function of the HZO thickness (right panel). Reproduced with permission.[73] Copyright 2022, AIP Publishing.

Amorphous Al$_2$O$_3$ is another typical material frequently used as a dielectric material in HfO$_2$-based FTJs with a TiN bottom electrode. Given Al$_2$O$_3$ has a similar coefficient of thermal expansion to TiN, it was expected to impose the comparable mechanical stress on the underlying HfO$_2$-based ferroelectric film, facilitating the attainment of an orthorhombic structure.[74] Moreover, it was reported that Al$_2$O$_3$ could suppress leakage and improve the breakdown characteristics.[75] Shekhawat *et al.* observed an ON/OFF current ratio of 78 in TiN/Hf$_{0.5}$Zr$_{0.5}$O$_2$ (HZO, 12 nm)/Al$_2$O$_3$ (1 nm)/Ge FTJs.[62] They found that using Ge substrates pretreated by in-situ ALD H$_2$-plasma effectively reduced the interfacial trap density state at the Ge-Al$_2$O$_3$ interface, and thus increased the ON/OFF current ratio. Compared to the wet etch treatment of the Ge substrate, the plasma treatment also led to fewer oxygen vacancies in $Al_2O_3$, and thus reduced background leakage current in FTJs.

Considering the ideal film thickness for the highest orthorhombic phase fraction in $HfO_2$-based ferroelectric film was often observed at ~10 nm,[76] Hoffmann *et al*. deposited 10 nm HZO on TiN coated p-type Si substrates by ALD, followed by reducing the HZO thickness using thermal atomic layer etching to retain the same amount of ferroelectric phase fraction.[73] A 2.9 nm-thick $Al_2O_3$ layer was subsequently grown on the remaining HZO layer which formed a ferroelectric/dielectric composite barrier of the final TiN/$Al_2O_3$/HZO/TiN FTJs (Figure 4c). The TER increased from 14 to 288 and then to 2469 when HZO thickness decreased from 10 nm to 7.4 nm and then to 6.2 nm, respectively. Simultaneously, the read voltage for maximum TER decreased from 2.3 V to 1.4 V and then to 1.3 V.

Note that in the ferroelectric/dielectric hybrid FTJs, the ON-state current could only tunnel through the dielectric when the polarization direction aligned with the tunneling direction.[64] However, the dielectric layer degraded dramatically during the memory operations, and thus limited endurance performance of FTJs due to dielectric breakdown caused by the electric field across the thin ultra-thin dielectric layer.[77] It was found that the dual $Al_2O_3$/$SiO_2$ dielectric layer could boost the performance of FTJ devices including the read disturbance, endurance characteristics and cell-to-cell tunnel current variation without degrading the ferroelectricity and the speed of polarization switching. For example, Min *et al*. fabricated TiN/$HfO_x$/$SiO_2$/Si single dielectric and TiN/$HfO_x$/$Al_2O_3$/$SiO_2$/Si dual dielectric FTJs.[78] Stable ferroelectricity was found in pure $HfO_x$ thin films, owing to the fast cooling process.[79] These improvements were ascribed to the decrease of the energy barrier for polarization switching due to the reduced grain size of ferroelectric $HfO_x$ induced by the additional $Al_2O_3$ layer on $SiO_2$.

## 2.2 Ferroelectric/Multiferroic Hybrid Junctions

### a. Electrical control of magnetism

The functionality of FTJs can be extended by incorporating ferromagnetic electrodes to promote the concept of multiferroic tunnel junctions (MFTJs). The integration of ferroelectricity and ferromagnetism in heterojunctions enables magnetoelectric coupling effects,[80–85] allowing for electrical control of spin polarization tunneling at the ferroelectric/magnetic interface, and thus the tunneling magnetoresistance (TMR).[86–90]

One limitation of spintronics is the substantial power consumption for magnetic writing processes. For example, a large external magnetic field is required to manipulate the spin degree of freedom in semiconductors through traditional techniques.[91,92] Zhuravlev *et al*. developed a method to switch the spin polarization of the current injected into a semiconductor by injecting spins from a dilute magnetic semiconductor through a ferroelectric tunnel barrier.[93] The results indicated that reversing the electric polarization of ferroelectric films significantly altered the spin polarization of the tunneling current, and thus enabled a dual-state electrical modulation of spin polarization.[93] A strong magnetoelectric coupling between the ferroelectric/magnetic interface is often attributed to orbital hybridization, charge carrier doping, lattice strain, or spin-dependent screening. This coupling is suggested to be responsible for the electrical control of spin polarization and magnetism by switching the direction of ferroelectric polarization.[83,94–96]

**b. Ferroelectric control of TER/TMR in MFTJs**

An MFTJ can be realized by integrating a magnetic tunneling junction (MTJ) and an FTJ by using ferromagnetic electrodes in an FTJ or using a ferroelectric barrier in an MTJ. As a result, TER and TMR effects co-exist in MFTJs. This hybrid junction offers the potential for advanced information processing and multiple resistance states, enabling high-density data storage.[20,88,89]

$La_{1-x}Sr_xMnO_3$ (LSMO) films are among the most extensively studied magnetic oxide materials in spintronic devices due to their rich physics and unconventional magneto-transport properties.[97,98] The electrical control of spin polarization was addressed experimentally in Fe/BTO/LSMO MFTJs, revealing a large and reversible dependence of the TMR on ferroelectric polarization direction.[81] In $Co/PbZr_{0.2}Ti_{0.8}O_3$ (PZT)/LSMO MFTJs, Pantel *et al*. found that if the ferroelectric polarization of PZT was switched from pointing towards Co to pointing towards LSMO then not only would the resistance of the junction be modified, but it would also have switched the TMR from inverse to normal (Figure 5a).[99] Moreover, the TMR could be switched back to its initial inverse sign by switching the ferroelectric polarization back. Similar results were obtained in LSMO/BTO/Ni MFTJs, where positive TMR was observed in the low resistance state (LRS) and negative TMR was observed in the high resistance state (HRS). This indicated that the reversal of ferroelectric polarization leads to the reversal of spin polarization.[100] In another study, Hambe *et al*. applied low-voltage pre-pulses to LSMO/BFO/LSMO MFTJs and found significant changes in the TMR (Figure 5b).[101] This phenomenon was ascribed to the modifications in the intricate band structure at the ferromagnetic-ferroelectric interface, resulting from ionic displacements associated with polarization reversal in the ferroelectric barrier.[102]

As a correlated electron oxide, the electronic and magnetic phases of LSMO are extremely susceptible to chemical composition, ferroelectric field, or strain.[31,32] Lee *et al*. demonstrated that the hole accumulation or depletion near the LSMO/ferroelectric interface in response to the polarization reversal could cause an insulator-metal transition. This transition created a large asymmetry, thereby leading to the TER effect.[31] Using interfacial- and element-sensitivity techniques of X-ray absorption spectroscopy (XAS) and X-ray magnetic circular dichroism (XMCD), Chi *et al*. provided direct experimental evidence of magnetoelectric coupling at the ferroelectric/magnetic interface in LSMO (x=0.5)/BTO heterostructures (Figure 5c).[103] They found that the LSMO/BTO interface was electron-doped when the polarization of BTO pointed toward the interface, and hole-doped when it pointed away from the interface. Accordingly, the main interfacial coupling of Ti-Mn was ferromagnetic when the polarization of BTO pointed toward the LSMO/BTO interface, while Ti−Mn was antiferromagnetic coupling when the polarization pointed away from the interface. This magnetic-phase transition induced by the reversal of ferroelectric polarization at the ferroelectric/magnetic interface not only enhanced the TER, but also modulated the spin filtering functionality of MFTJs, thereby enabling the enhancement of the TMR effect.[103]

Composite barriers are also employed in MFTJs to improve device functionality. In MFTJs with a STO/BTO dielectric/ferroelectric barrier, the dielectric STO layer between ferromagnetic electrodes enhanced the TER ratio by two orders of magnitude.[105] This enhancement was ascribed to the increase in the asymmetry of the electrostatic modulation on the barrier profile. Meanwhile, an improved TMR ratio was observed, which was explained by the enhanced spin polarization at the ferromagnetic/dielectric interface.[105] The TER ratio was also enhanced by two orders of magnitude by utilizing a bilayer tunneling barrier BTO/La$_{0.5}$Ca$_{0.5}$MnO$_3$ (LCMO).[32,106] This improvement was likely related to the ferroelectrically controlled metallic–insulating phase transition in LCMO, while the ferroelectrically controlled TMR was owed to the strong magnetoelectric coupling at the BTO/LCMO interface.

Metallic–insulating phase transitions can occur not only in magnetic manganese oxides, but also in ferroelectrics.[84,107] First-principle calculations on Co/(TiO$_2$-BaO)$_5$/Co MFTJs revealed that the interfacial Ba-O displacement was much larger than the ferroelectric displacement in the BTO bulk when the polarization of BTO pointed to the Co-TiO$_2$ interface, which moved the energy band of BTO to a lower energy and thus led to a metallic BTO. For right-polarized Co/(TiO$_2$-BaO)$_5$/Co MFTJs (*i.e.,* the case of the polarization pointed to the BaO-Co interface), the interface ferroelectric displacement was not significantly higher than that of the bulk, implying a weak influence of interface on the BTO band and a normal insulating state for the right-polarized BTO.[108] Similar results were obtained in Co$_2$MnSi/BTO/SRO MFTJs, where the displacement at the MnSi-TiO$_2$ interface was giant (small) for the right (left)-polarized state (Figure 5d).[104] This corresponded to a ferroelectric barrier height close to zero (a sizable barrier height), leading to an optimistic TER ratio of up to $1.1 \times 10^3$.

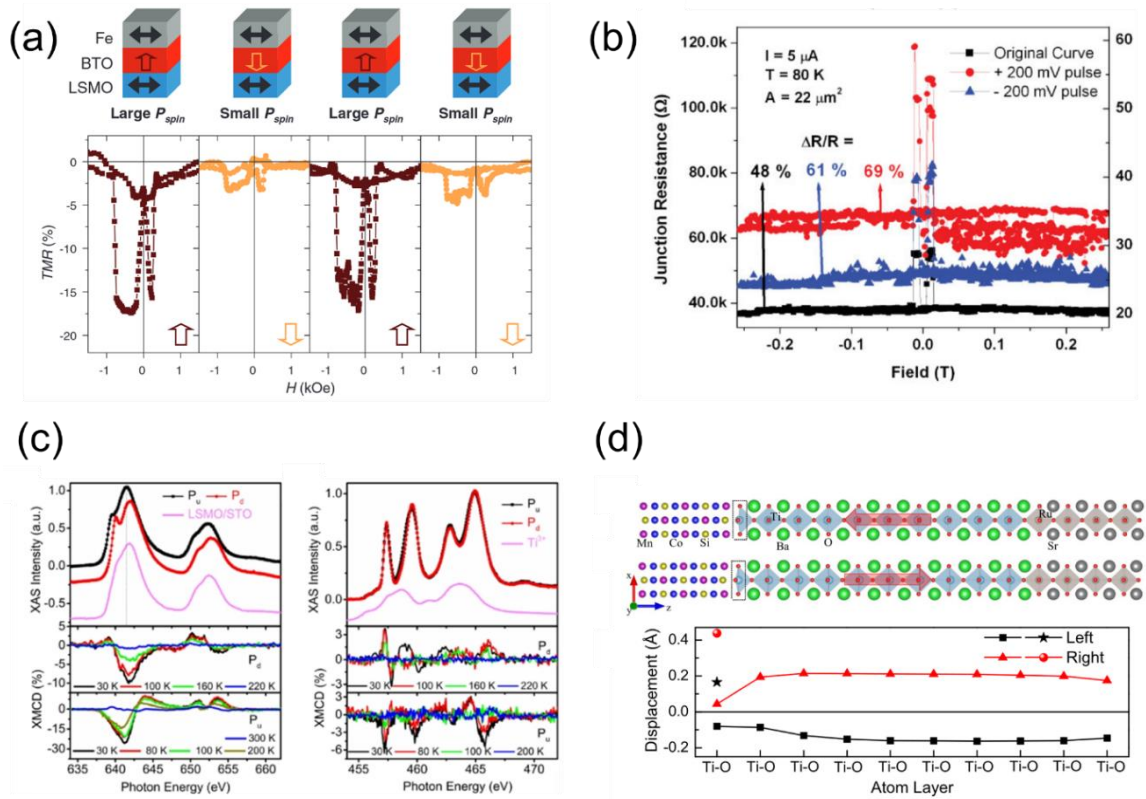

**Figure 5.** a) TMR curves of Fe/BTO/LSMO MFTJs recorded after poling the ferroelectric up, down, up, and down, sequentially. Reproduced with permission.[81] Copyright 2010, AAAS. b) TMR of LSMO/BFO/LSMO showing the effect of positive and negative bias pulses. Reproduced with permission.[101] Copyright 2010, Wiley-VCH Verlag GmbH & Co. KGaA. c) Polarization dependence of XAS and XMCD at Mn and Ti L$_{2,3}$ edges showing magnetoelectric coupling at LSMO/BTO interface for different polarization directions of BTO. Reproduced with permission.[103] Copyright 2021, American Chemical Society. d) Relaxed structures of the Co$_2$MnSi/BTO/SrRuO$_3$ MFTJs with a MnSi-TiO$_2$ termination for different polarization directions of BTO. The relative displacement along the z direction for each TiO$_2$ layer is shown in the bottom panel. Reproduced with permission.[104] Copyright 2022, American Physical Society.

## 2.3 Ferroelectric/2D hybrid junctions

2D materials refer to material systems that have layered structures that are only one or a few atom layers thick, typically up to a few nanometers. Due to their intrinsic thinness and high surface-to-volume ratio, these materials exhibit novel physical properties such as unique band structures, quantum confinement, 2D excitons, commensurate-incommensurate transition, exceptional mechanical flexibility, high optical transparency, and high carrier mobility, which can enhance device efficiency.[109,110] One notable type of 2D materials are the so-called van der Waals (vdW) materials. Their atoms within the same layer are interconnected through strong chemical bonds, while atoms between different layers are connected by the relatively weaker van der Waals force.[111] Interest in 2D vdW materials has been extremely strong since the report of graphene in 2004.[112] Sustained efforts in 2D materials have led to the emergence of several 2D vdW ferroelectrics, such as $CuInP_2S_6$ (CIPS),[113–115] $In_2Se_3$,[116–118] and ferromagnetics such as $Fe_nGeTe_2$.[119,120]

As a result, 2D vdW materials, including graphene, $MoS_2$, WeS, have found wide applications in transistors, energy devices, photovoltaic devices, memory elements, flexible electronics, etc. In respect of ferroelectric based memory devices, hybrid junctions combining conventional ferroelectric thin films with 2D materials have shown distinct characteristics and innovative applications driven by the 2D-FE interface couplings. Notably, the switchable polarization of ferroelectrics in these hybrid junctions offers an additional degree of device manipulation. Examples include 2D tunnel junctions for memristive devices[121,122] and 2D ferroelectric field-effect transistors (FeFET).[123] The unique properties of 2D materials are expected to provide a new and more functional platform for next-generation electronics.

In this section, we discuss 2D ferroelectric hybrid tunnel junctions, including conventional perovskite FTJs with a 2D vdW semiconductor electrode (2D/FE FTJs) and 2D vdW FTJs in which the ferroelectric barriers are 2D vdW materials.

### a. 2D/FE FTJs

Ferroelectric resistive switching behavior in 2D/FE FTJs is driven by the interface modulation of contact barriers and depletion width through tuning the polarization of the ferroelectric layer.[25,124] Thus, the interface properties play an important role in determining FTJ performance. Several ways have been reported to effectively engineer the interface, such as the type and thickness of 2D material, introduction of molecular layers, etc. Both graphene/FE and $MoS_2$/FE heterojunctions and their ferroelectric resistive switching behaviors in FTJs have been intensively studied.[121]

Graphene has been a popular top electrode for FTJs due to its near zero band gap and high conductivity.[125,126] A very large modulation on the device current (ON/OFF ratio of $10^5$) had been achieved by adjusting the gate voltage to control the graphene–silicon Schottky barrier.[127] As mentioned in Section 2.1, when an additional dielectric layer was introduced in the FTJ to form a ferroelectric/dielectric barrier, it acted as a switch that further tuned the barrier height upon polarization switching of the ferroelectric layer. This adjustment occurred through the changing electrostatic potential caused by the incompletely screened interface charge.[36] Using this idea, Gruverman's team first introduced molecular layers at the graphene/BTO interface.[122] Molecular layers could provide multiple tuning degrees, such as molecule size, shape and dipole features. In this study, the PMMA/graphene layer was treated in a water and ammonium hydroxide solution respectively before being transferred to the BTO surface, to introduce either a water or ammonia molecular layer between graphene and BTO. It was found that the FTJ with an ammonia molecular layer at the graphene/BTO interface showed much more stable polarization (up to ~4 ×$10^3$ switching cycles) compared to the FTJ with a water layer, while also exhibiting a high OFF/ON resistant ratio up to >3800 in the zero-bias limit and ~ 6000 under ± 0.2 V bias range (Figure 6a). This TER was much higher than the BTO FTJ without graphene and a molecular layer, of which the TER magnitude was merely about 7.[128] As illustrated in Figure 6b, the stable polarization and enhanced TER could be attributed to the dipole moment of ammonia molecules which switched along with the BTO polarization. On the other hand, the dipole moment of water molecules had a strong in-plane component, and thus led to unstable polarization in graphene/$H_2O$/BTO.[122]

Molybdenum disulfide ($MoS_2$) is another popular 2D semiconductor used in FTJs. Monolayer $MoS_2$ has a bandgap of 1.8 eV,[129,130] and it can contribute to an asymmetric polarization switching behavior in $MoS_2$/BTO junctions (Figure 6c). In $MoS_2$/BTO/SRO FTJs, a giant TER was also observed with ON/OFF resistance approaching $10^4$ (Figure 6d~e). This value was nearly 50 times that of FTJs with metal electrodes[131] and this dramatic enhancement could be attributed to the barrier height change at the $MoS_2$/BTO interface upon polarization switching, which caused reversible accumulation/depletion of carriers in $MoS_2$.[132]

On the other hand, the resistive switching diode performance does not show significant improvement in graphene/BFO over metal/FE diodes, possibly owing to the poor polarization stability at the graphene/BFO heterointerface. For example, the ON/OFF ratio reading under 1 V *(i.e.,* approximately 200) of bilayer graphene/BFO was close to the switchable diode with metal electrodes.[121] Nevertheless, the authors demonstrated that the thickness of 2D materials played a crucial role in determining the resistive switching behavior and TER, as it influenced the contact barriers and depletion layer and thus the work function. This made it possible to further tune the resistive switching and TER through the thickness control of 2D semiconductor electrodes. It was found that less graphene layers (*e.g.,* bilayer) contributed to a higher ON/OFF resistance ratio compared to the multilayers (up to 14 layers), while maximum ON/OFF ratio (*i.e.,* 190) was achieved in MoS2/BFO when $MoS_2$ had a thickness of ~4 nm (*i.e.,* 5-layer).[121] Such thickness were also demonstrated through theoretical calculations for the previously mentioned molecular layer structured graphene/FE FTJ. It was found that the TER ratio increased exponentially with the ammonia molecule layer thickness.[122]

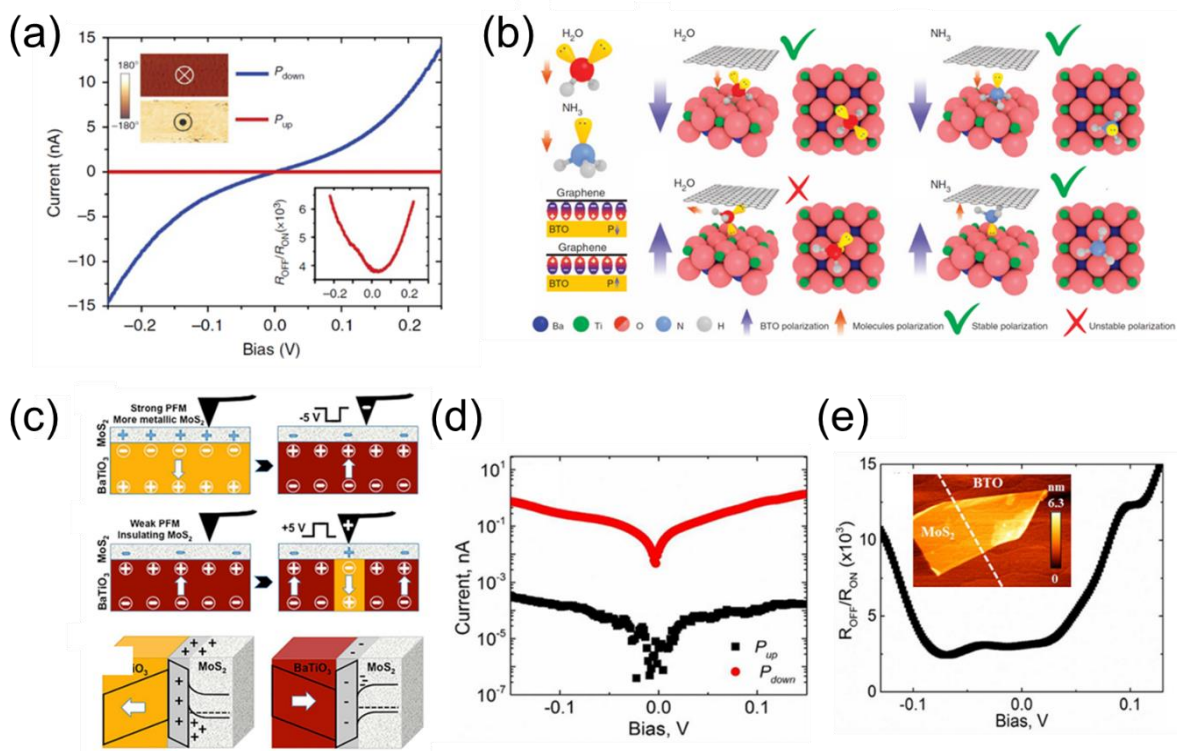

Figure 6. a) I-V curves upon poling and corresponding $R_{OFF}/R_{ON}$ resistant ratio curve (inset) of the Graphene/BTO/LSMO junction with $NH_3$ molecular layer; b) Illustration of polarization retention differences between Graphene/BTO junctions with water and ammonia molecules. Reproduced with permission.[122] Copyright 2014, Springer Nature Limited. c) Asymmetric switching behavior of $MoS_2$/BTO/SRO junction; d) I-V curves upon poling and e) $R_{OFF}/R_{ON}$ resistant ratio curve of $MoS_2$/BTO/SRO junction. Reproduced with permission.[132] Copyright 2017, American Chemical Society.

## b. 2D vdW FTJs

Unlike conventional perovskite ferroelectrics (*e.g.,* BTO) whose ferroelectricity disappears below a critical thickness,[12] ferroelectricity in 2D vdW ferroelectrics, such as CIPS, $\alpha$-In$_2$Se$_3$, SnTe, and 1T-MoTe$_2$, survives at few-layer thicknesses, even towards the monolayer limit.[116,119,133] In addition, the dangling-bond-free surfaces and minimal surface of 2D materials allow for the construction of ideal interfaces with dissimilar materials and good compatibility with industrial silicon electronics.[134–136] These advantages are beneficial to the miniaturization and performance improvement of FTJs by using 2D ferroelectrics as the tunneling barriers.

To achieve a large TER, a significant modulation of the barrier width and/or height is necessary. TER values beyond $10^6$ had been realized in perovskite-based FTJs with a Nb-doped SrTiO$_3$ (NSTO) semiconductor electrode, owing to the parallel modulation of barrier width and height by the ferroelectric field effect.[137,138] Nevertheless, such perovskite structured devices are not compatible with existing electronics technology.

Recently, Wu *et al*. fabricated FTJs based on a 2D vdW heterostructure, using 4 nm-thick CIPS as the ferroelectric tunneling barrier layer and chromium and monolayer graphene as asymmetric electrodes.[139] They reported a giant TER of above $10^7$ in these FTJs, attributed to a giant barrier height modulation of approximately 1 eV, resulting from the polarization reversal induced Fermi level shift in the semi-metallic graphene (Figure 7a). This substantial ferroelectric modulation of the barrier height in 2D Cr/CIPS/graphene FTJs was also reported by Wang *et al*, who obtained a giant TER of $10^9$.[140] Moreover, the TER could be further enhanced to $10^{10}$ by inserting a monolayer MoS$_2$ between CIPS and graphene. MoS$_2$ acted as an n-type semiconductor at the ON state while it became more intrinsic (*i.e.,* more insulating) at the OFF state due to the ferroelectric field effect.[132,140] As a result, little change was observed at the ON-state current, while the OFF-state current was further suppressed with the insertion of monolayer MoS$_2$, leading to enhanced TER.

A special 2D vdW ferroelectric material, $\alpha$-In$_2$Se$_3$ has attracted great attention due to its low coercive electric field, high Curie temperature, as well as intercorrelated out-of-plane and in-plane ferroelectricity at monolayer thickness.[118,141–143] All-2D vdW hexagonal boron nitride (h-BN)/$\alpha$-In$_2$Se$_3$/multilayer graphene (MLG) FTJs were fabricated by Tang *et al*.[144] In this configuration, few-layer h-BN acted as the tunnel barrier, and 8 nm-thick $\alpha$-In$_2$Se$_3$ served as the core ferroelectric material, and relatively thick MLG functioned as the bottom electrode, favorably for improving the electrical transport properties (Figure 7b). A large ON/OFF ratio of $10^4$ was achieved by switching the ferroelectric polarization in $\alpha$-In$_2$Se$_3$ at room temperature accompanied with a retention time over $10^4$ s. It was suggested that the Fermi level shift on the $\alpha$-In$_2$Se$_3$ surface in response to the polarization reversal contributed to the large resistance change.[144] Theoretical investigations on $Fe_3GaTe_2/\alpha-In_2Se_3/Fe_3GaTe_2$ vdW MFTJs predicted a giant TMR over 10,000% by switching the magnetic alignments of $Fe_3GaTe_2$, and TER exceeding 300% by controlling the ferroelectric configurations of bilayer $\alpha-In_2Se_3$.[145,146] Furthermore, both the TMR and TER ratios of the MFTJs could be greatly enhanced by introducing interface asymmetry and inserting monolayer of h-BN at one interface.[145]

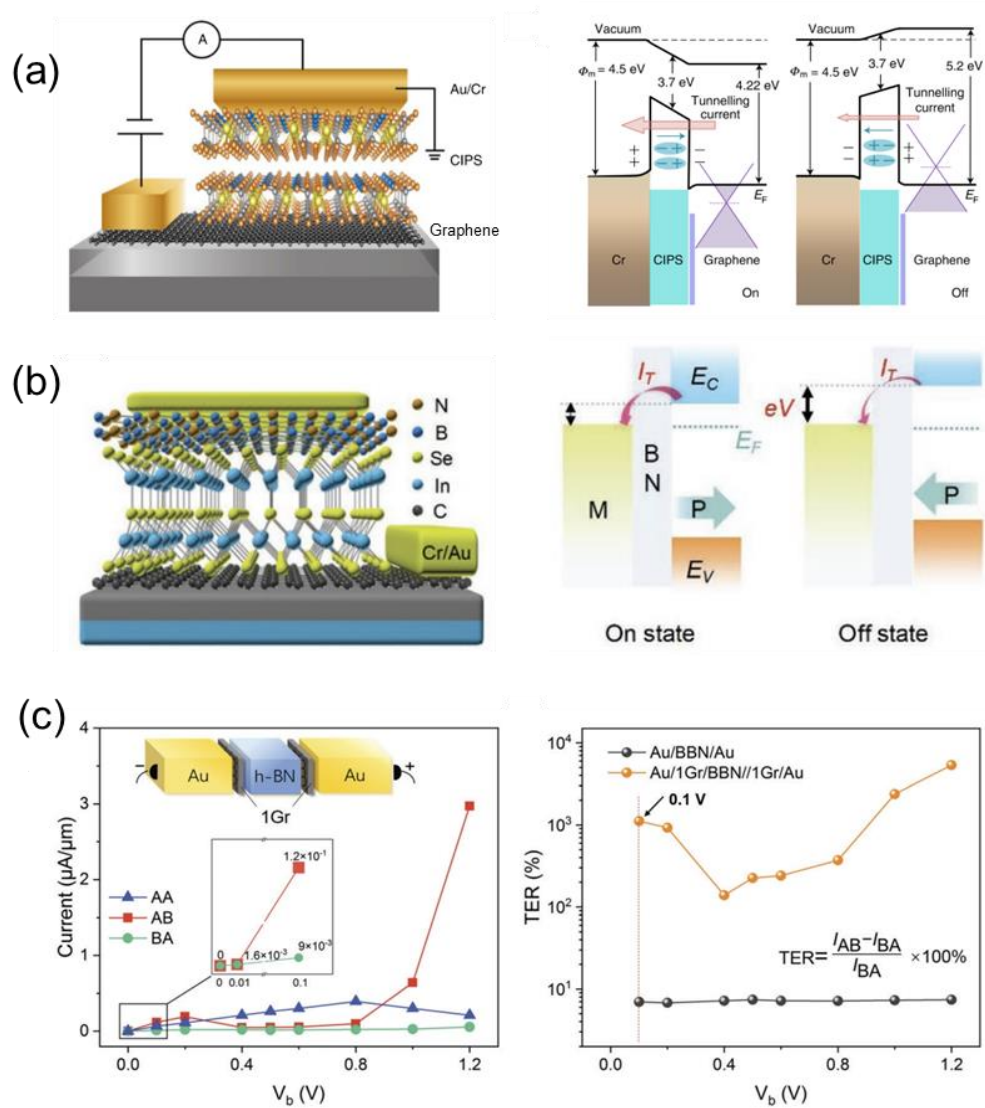

**Figure 7.** a) Schematic of the Cr/CIPS/graphene FTJ on the $SiO_2$/Si substrate (left panel) and band diagrams for the On and Off states of the vdW FTJ operation (right panel). Reproduced with permission.[139] Copyright 2020, published by Springer Nature. b) Schematic illustration of the vdW h-BN/$\alpha$-$In_2Se_3$/MLG FTJ (left panel) and band diagram of the tunneling process between the $\alpha$-$In_2Se_3$ surface and the top metal electrode showing the working mechanism of the vdW FTJ (right panel). Reproduced with permission.[144] Copyright 2022, Wiley-VCH GmbH. c) I–V outputs (left panel) and the calculated TER (right panel) of the bilayer BN ferroelectric tunnel junctions with monolayer graphene (1Gr) intercalation (Au/1Gr/bilayer h-BN/1Gr/Au). Reproduced with permission.[147] Copyright 2022, The Royal Society of Chemistry.

Unlike non-stacking vdW ferroelectrics (*e.g.,* CIPS and α-In$_2$Se$_3$) whose ferroelectricity arises from intrinsic lattice symmetry breaking, some vdW materials with centrosymmetric lattice structures can exhibit ferroelectricity by stacking. These group of materials are known as sliding ferroelectrics, with h-BN being a notable example.[148,149] In bilayer h-BN, when the N atoms in the upper layer are directly above the center of the hexagonal rings in the lower layer, and the B atoms in the upper layer are directly above N atoms in the lower layer, a metastable AB or BA configuration forms and creates an asymmetry between the two layers. This leads to a redistribution of charge, with net charge transferring from the upper layer to the lower layer, resulting in a vertical polarization directed upwards (Figure 7c).[150] The polarization of the system can be reversed by interlayer sliding of the B-N bond length. For example, the N atoms in the upper layer can be directly aligned above the B atoms in the lower layer by moving the upper layer. Yang *et al.* demonstrated the transport behavior in vdW sliding FTJs consisting of a bilayer h-BN as an exceedingly thin tunneling barrier sandwiched between Au electrodes.[147] Their research, which leveraged density functional theory (DFT) in conjunction with the nonequilibrium Green's function approach to analyze the transport characteristics, revealed a minimal TER due to the unexpected zero polarization in the h-BN. This effect resulted from the strong Au/h-BN interfacial electric field when bilayer h-BN was directly deposited on Au electrodes. When monolayer graphene was interposed between the Au electrode and the h-BN, a large TER of approximately 10,000% was observed (Figure 7c). The key factor contributing to this dramatic increase in TER was the ability of graphene to reduce the intense hybridization between the Au and the adjacent h-BN, thereby recovering the ferroelectricity of the bilayer h-BN.[147]

**2.4 Ferroelectric/Superconductor Hybrid Junctions**

Electrical control of superconductivity remains a topic of significant interest in both scientific research and technological applications.[151] The external field is capable of altering the carrier concentration and resistance of superconductors, which plays a key role in the development of superconducting field-effect transistors (SuFETs) and superconducting quantum interference devices (SQUIDs).[152–154] For example, the ferroelectric field effect in superconductor (SC)/ferroelectric (FE) junctions had been reported to trigger the accumulation or depletion of carriers in superconductors, and thus enhanced or suppressed the superconducting critical temperature ($T_c$).[155] A significant change in both critical current and the device normal resistance was also observed in a SQUID as a result of polarization reversal.[156]

Cuprate superconductors, such as Bi-Sr-Ca-Cu$_2$-O$_X$ (BSCCO) and YBa$_2$Cu$_3$O$_7$ (YBCO) are recognized for their high-temperature superconductivity with a T$_c$ of about 110 K and > 90 K at ambient pressure, respectively.[157,158] Specifically, YBCO possesses a perovskite crystallographic structure characterized by lattice parameters of a = 3.82 Å, b = 3.89 Å, and c = 11.67 Å.[159] Notably, its in-plane lattice parameter closely aligns with that of several perovskite ferroelectrics, including BTO (a=3.993 Å) and BFO (a=3.958 Å). This close match in lattice parameters renders YBCO an exceptional superconductor material for pairing with ferroelectrics in heterostructure junctions, facilitating the integration of diverse functionalities in such hybrid junction systems. In YBCO, its crystallographic structure and transport behavior (*e.g.,* conductivity) are strongly influenced by the oxygen stoichiometry.[160,161] For example, as the oxygen level decreased in YBa$_2$Cu$_3$O$_x$ from x=7 to $x \leq 6.3$, YBCO underwent a structural transition from orthorhombic to tetragonal symmetry. During this transition, the resistivity increased by 9 times,[160] while the T$_c$ dropped to less than 10 K.[162] In addition, the T$_c$ and resistance are also strongly dependent on the BFO polarization direction (Figure 8a).[163]

Ravikant *et al*. investigated the tunneling current in LSMO/PZT/BSCCO magnetic-ferroelectric-superconducting heterostructures.[164] The heterostructures showed the electric field control of TER with distinct resistive states for different polarization orientations even at room temperature, and a TER effect that was 1~2 orders higher at 78 K (below the T$_c$ of BSCCO). Also, the effect of the perpendicular magnetic field on the TER was significant at room temperature but almost disappeared at 78 K due to the Meissner effect of the superconducting electrode.[164] In order to explore the interplay between superconductivity and TER, the Villegas group fabricated MoSi/BFO/YBCO (SC/FE/SC) hybrid junctions.[33] Reversible resistance switching was observed with the voltage-induced reversal of the ferroelectric polarization. However, because the behavior was similar to those in MoSi/YBCO junctions, and in junctions with a non-ferroelectric STO interlayer, ferroelectricity did not play a central role, in contrast to the conventional FTJs where tunneling was dominated by charge migration-induced barrier height change at the interface.[25]

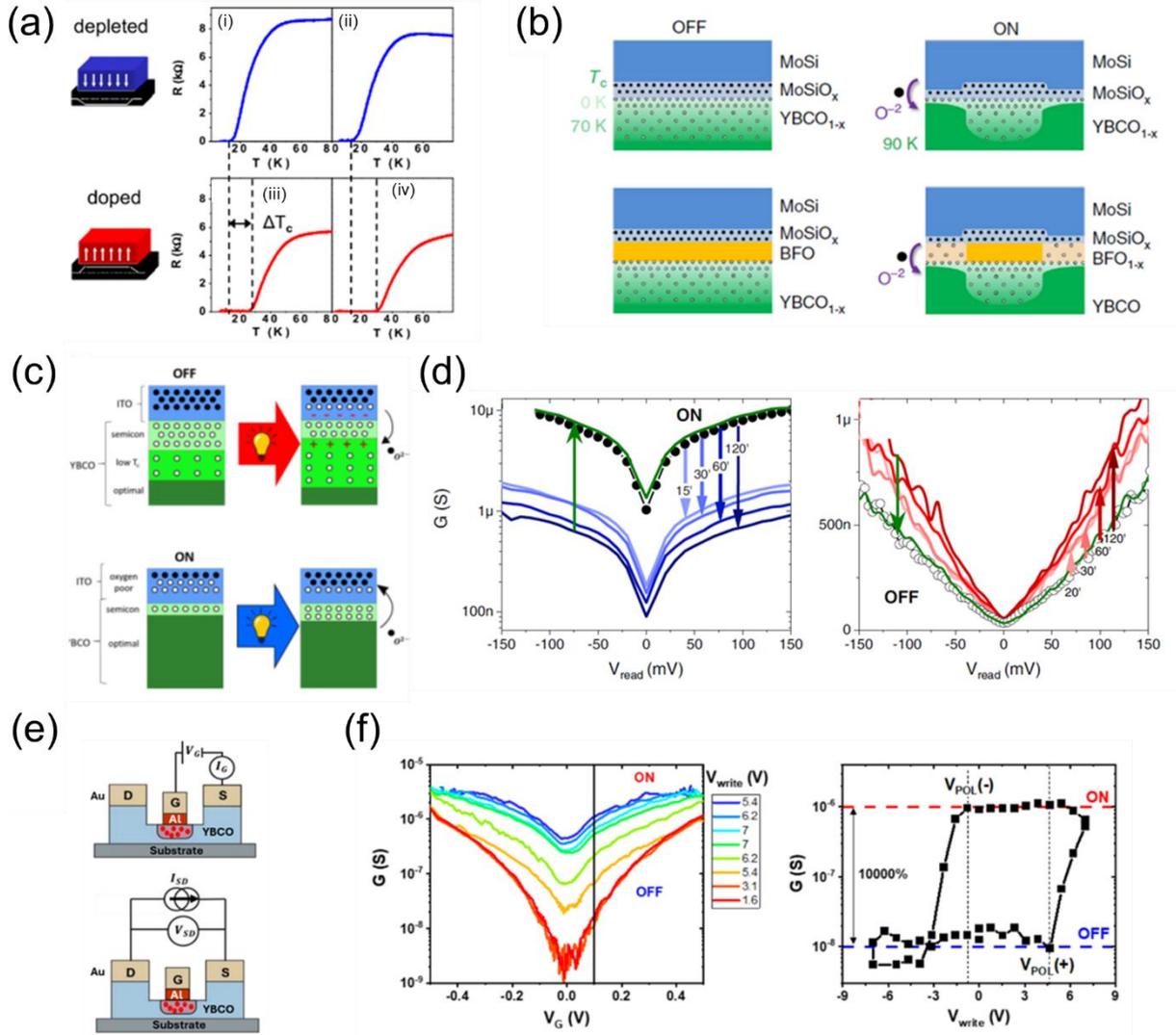

**Figure 8.** a) Resistance versus temperature for a BFO (30 nm)/YBCO (4 u.c.)/ PrBa$_2$Cu$_3$O$_7$ (PBCO, 2.4 nm)//STO heterostructure in the "as-grown" state (i), and after subsequently reversing the ferroelectric polarization outwards (ii), towards (iii), and outwards (iv) the YBCO layer. Reproduced with permission.[163] Copyright 2013, American Institute of Physics. b) Electro-resistance model based on electrochemistry. Reproduced under the terms of the Creative Commons CC BY license.[33] Copyright 2020, The Author(s), published by Springer Nature. c) Schematic of the ITO/YBCO interface during optical switching from OFF to ON state (top panel) and from ON to OFF state (bottom panel); [165] d) Differential conductance as a function of $V_{read}$ before and after illumination as a function of illumination period. Reproduced under the terms of the Creative Commons CC BY license.[165] Copyright 2023, The Author(s), published by Springer Nature. e) Schematic representation of the measuring setup used for the transport measurement of Al/YBCO; [166] f) (left panel) Differential conductance as a function of reading voltage for a YBCO/Al device with YBCO and Al thicknesses of 50 nm and 10 nm respectively. (right panel) Differential conductance as a function of $V_{write}$ measured under a fixed reading voltage of 100 mV. Reproduced with permission.[166] Copyright 2024, IOP Publishing Ltd.

In superconductor hybrid junctions, the resistance switching could occur through reversible electrochemical redox reactions. Cuprates have a strong tendency to lose oxygen, primarily due to the high reduction potential of $Cu^{+3}$ and an oxygen depletion layer would form within.[33,165,166] Consequently, this could lead to a less conductive (*e.g.,* semiconductive or low $T_c$) interfacial layer, as illustrated in Figure 8a.[165] In MoSi/YBCO and MoSi/BFO/YBCO systems, MoSi exhibited a strong oxidation potential, and it was anticipated that MoSi would undergo spontaneously oxidation when deposited on top of the YBCO or BFO surface.[33] The multiferroic BFO is also prone to losing oxygen and forming oxygen vacancies during deposition. Upon application of an external electric field, oxygen exchange would occur at the junction interface, leading to changes in oxygen stoichiometry within YBCO. This facilitated the realization of resistive switching between OFF and ON states, often with several intermediate states and consequently, altering its resistivity (Figure 8b).[33,165] Such redox reaction dominated resistive switching in SC/FE and SC/SC junctions and it was found that the superconducting energy gap significantly enhanced TER effects up to 3000% larger than conventional electrodes.[33]

Due to the interfacial redox reaction, the ON/OFF state could be observed without a ferroelectric barrier when superconductor materials are involved as the electrode.[33,165,166] It had been observed in various combinations, such as YBCO/ITO,[165] YBCO/Al,[166] where ITO and Al all exhibited potential for oxidation (Figure 8c-f). Nevertheless, such switching behavior was largely attenuated when the paired materials are replaced by non-oxidizing materials.[165]

## 2. Ferroelectric Resonant Tunneling Diodes (FeRTDs)

So far, we have described the tunneling across hybrid junctions where the combination of the ferroelectric with the functional layer effectively acts as "one" material system. In this section we introduce FeRTDs - now the functionality stems from different layers interacting with each other to form a potential well sandwiched inside a quantum-well (QW) structure. Note that we do not strictly differentiate between FeRTDs and FTJs that have resonant tunneling effect in this perspective.

We start with an introduction to resonant tunneling. Resonant tunneling is a quantum-mechanical effect that allows electrons to tunnel through a quantum-well structure with an increased probability when the electron injection energy aligns with a discrete quantized energy level within the well.[167–169] It is often accompanied by the negative differential resistance (NDR) phenomenon, which represents a decrease of current due to the alignment and subsequent misalignment of the injection and QW energy levels under increasing bias.[170–172] Under the confinement of the two potential barriers, a quantized discrete set of energy states are formed in the potential well. Applying a positive voltage V to the left electrode will raise energy of this electrode by eV (Figure 9a). Thus, the quantum-mechanical effect of resonant tunneling occurs when the electron injection energy of the left electrode aligns with a discrete energy level (Figure 9b). As the bias continues to increase, the bottom of the conduction band of the left electrode aligns with a discrete energy level at the resonant voltage $V_R$, resulting in a peak in the current, $I_P$ (Figure 9c). Increased bias past this resonant bias is often accompanied by the phenomenon of NDR (Figure 9d), *i.e.,* a reduction in current due to the electrons no longer having an appropriate discrete well energy level. Upon further increasing bias it is possible to observe a second $I_p$ current when electron injection energy aligns with a higher QW energy level.

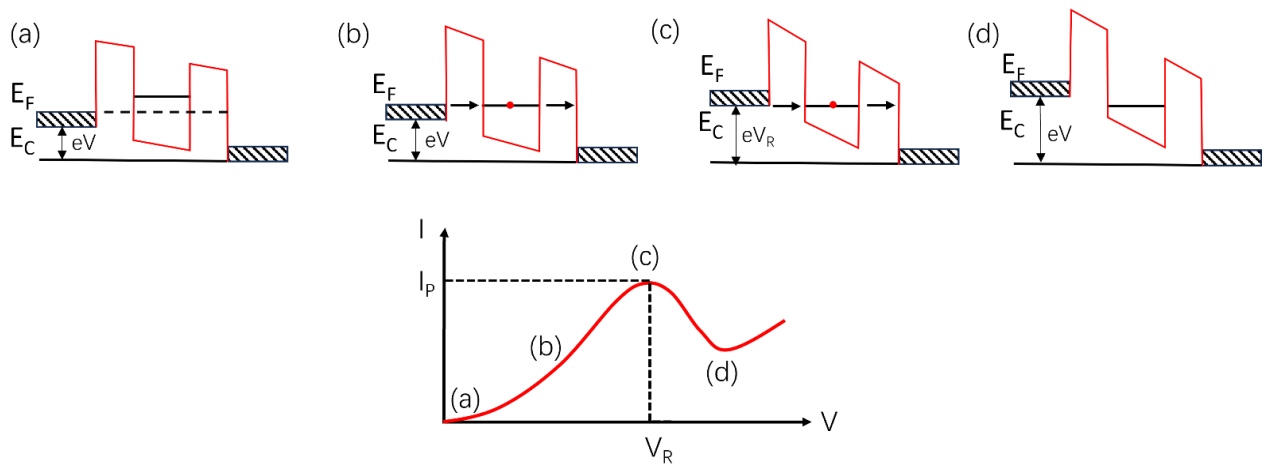

**Figure 9. Schematic band structure and I-V curve of a RTD under various biases. a) An initial positive bias is applied at the left electrode. b,c) Resonant tunneling through a discrete energy level occurs when the discrete energy level drops between the Fermi level and bottom of conduction band of the left electrode. e) I-V curve illustrating the occurrence of resonant tunneling followed by NDR.**

Predictions of resonant tunneling through double barriers trace back as early as 1951 by Bohm and were extended to multiple barriers (superlattices) by Tsu and Esaki in 1973.[173,174] The following year, experimental realization with I-V characteristics of resonant tunneling and NDR were observed by Chang *et al*.[175] As mentioned previously, recent advances in thin film fabrication techniques have enabled the high-quality crystalline structure of oxides, coupled with the construction of two-dimensional heterostructures featuring barrier layers just nanometers thick. Then the combination of ferroelectricity and QW structures at the nanoscale gave rise to the concept of ferroelectric resonant tunneling diodes (FeRTDs). The classic FeRTDs are a class of hybrid junctions where the typical ferroelectric single layer in an FTJ is replaced with a QW multi-layer structure. This structure consists of a potential well which acts as the functional layer sandwiched between two potential barriers, in which the barrier or well are ultrathin ferroelectric films. The polarization direction of the ferroelectric layer plays a critical role in shaping the tunneling profile of FeRTDs, and thus leads to significant shifts in both the peak voltage and peak current upon reversal of the ferroelectric polarization. This control over polarization enables the precise modulation of resonant tunneling and associated NDR behaviors. Consequently, FeRTDs offers opportunities for advanced electronics with tunable and highly responsive functionalities.

**2.5 Types of FeRTDs**

Various types of FeRTDs have been realized, making use of the wide array of properties in FE materials. These include harnessing the polarization within the well, using a ferroelectric as the barriers, or even making use of the conductivity properties of FE domain walls.

**a. Ferroelectric potential well**

Early studies on FeRTDs primarily employed the ferroelectric layer as the potential well. Theoretical investigations from an early stage indicated that the spin polarized transport could be tuned by switching the ferroelectric polarization.[93] Li *et al*. proposed a spin switch based on a FeRTD with resonant tunneling and NDR effects.[176] It showed that reversal of the ferroelectric polarization of the potential well led to a noticeable change in the spin polarization. At a critical bias, a highly spin polarized current or evenly spin mixed current could be obtained, depending on the polarization orientation of the ferroelectric layer in the QW.[176] The same authors had suggested that this phenomenon occurred due to asymmetry from nonequilibrium transport in either polarization direction.[177] However, it should be noted that these investigations were based on Li-doped ZnO. While it had been considered as a candidate ferroelectric back to the late 1990s,[178] doubts have been raised regarding its true ferroelectric nature, due to indications that its electrical hysteresis stems from mobile defect charge effects, rather than the reorientation of Zn-O bonds.[179] Later Du *et al*. experimentally demonstrated the ferroelectric modulation of resonant tunneling and NDR using the prototypical ferroelectric BTO as the potential well.[180] In their study, modulation of NDR was achieved by ferroelectric switching. For example, when the polarization was directed towards the collector, the depolarization field would point towards the emitter. This lowered the quantized energy levels in the BTO potential well layer, necessitating a lower resonant bias for NDR. Conversely, if the polarization was pointed to the emitter, the extra depolarization field raised the quantized energy levels. Thus, a higher resonant bias was required for NDR in the opposite polarization direction.[180]

**b. Ferroelectric barrier**

Although some inroads into FeRTDs have been made, the pursuit of better performance in ferroelectric tunnel heterostructures have led to preliminary studies where ferroelectrics are used as the barriers and the well layer is a dielectric.[181,182] This concept was touched upon in a study on a SRO/BTO/SRO FTJ from Su *et al*.[183] They suggested that the substitution of one $TiO_2$ layer with $SnO_2$ could be considered as the insertion of a one unit cell $BaSnO_3$ layer, thus creating a BTO/$BaSnO_3$/BTO QW heterostructure. From first-principles calculations, the authors argued that dependent on the position of the $BaSnO_3$ and the direction of polarization, the transition from a direct tunneling regime to resonant tunneling induced a giant TER effect. Another study in physical modelling showed that a HZO/$Ta_2O_5$/HZO QW with ferroelectric barriers exhibited NDR and a TER enhanced by several orders of magnitude due to resonant tunneling.[184]

As the reader may have noticed, theoretical studies in FeRTDs vastly outweigh the successful experimental demonstrations, pointing to the practical difficulties of fabricating the additional layers compared to a conventional FTJ. On the other hand, up to the past decade, the study of room-temperature resonant tunneling and the associated NDR has mainly focused on silicon-based and III-V group compound semiconductors.[185–187] There had been a lull in experimental fabrication of an FeRTD until very recently. In 2022, we resolved this longstanding problem by fabricating a FeRTD with a QW heterostructure comprised of BTO/SRO/BTO.[188] With SRO, a metallic material as the QW layer, our work reported the largest OFF/ON resistance ratios values in tunneling devices that utilize BTO as the barriers. Importantly, the results showed the hallmarks of resonant tunneling, followed by NDR at room temperature. Significant modulation of this behavior was achieved owing to charge distribution from polarization switching. Additionally, the observation of an inverse relationship between the current peak and temperature, and NDR in only the positive bias direction was consistent with previous work.[189] Through DFT calculations, the occurrence of NDR was attributed to the growth of orthorhombic SRO and the electron-electron correlations between its Ru-$t_{2g}$ and Ru-$e_g$ states.[188] The fabrication of such a room temperature FeRTD with good retention and cyclic endurance provides exciting prospects to the future of ferroelectric hybrid junctions.

**c. (Head-to-head) domain wall based FeRTDs**

Domain walls in ferroelectrics are considered promising active elements for next-generation non-volatile memory, logic gates and energy-harvesting devices due to their unique characteristics, including enhanced or reduced electrical conductivity, reversible creation, movement and destruction under external fields, [190–192] Unlike normal QWs consisting of three-layer composite films, a V-shaped electrostatic potential profile across a single ferroelectric layer resulting from the head-to-head domain wall can form a barrier/well/barrier QW structure. This occurs as the conduction band was pulled down due to the presence of positive polarization charges near the domain wall. Experimental quantum oscillations of conductance indicating resonant tunneling were observed in LSMO/BTO/LSMO MFTJs (Figure 10a).[192] The measurements showed that there was a head-to-head domain wall within the ultrathin BTO film induced by oxygen vacancies. Moreover, the head-to-head domain wall could be removed by switching the ferroelectric polarization of BTO.

Inspired by the above work, Li *et al*. investigated the role of a head-to-head domain wall acting as a QW.[193] Through first-principles DFT, the authors demonstrated that in a LSMO/BTO/LSMO heterostructure, head-to-head domain walls could be stabilized by polar boundaries, *i.e.,* the $(La_{0.5}Sr_{0.5}O)^{0.5+}/(TiO)^0$ terminated interfaces. As a result, an acting QW structure was created due to the V-shaped electrostatic potential profile across the BTO. Confined to the domain wall, a 2DEG was formed which facilitates resonant tunneling through localized electronic states. It was reported that a pinned dipole layer which was interface dependent and not switchable could be formed at the metal/ferroelectric interface due to the interface effect.[194] For metal/pinned-dipole-layer/ferroelectric/metal hybrid tunnel junctions, a head-to-head domain was formed when the polarizations of the pinned-dipole-layer and ferroelectric point to each other. Therefore, resonant tunneling occurred as the bottom of the V-shaped well was near the Fermi level.[182]

In addition, first-principles calculations and quantum-mechanical tunneling simulations showed that resonant tunneling may also occur in FTJs with a ferroelectric/non-polar dielectric composite barrier.[181,182,195] For example, when the polarization of the ferroelectric pointed to the adjacent dielectric layer that has a small dielectric constant, the electric field in the dielectric will be strong and opposite to the direction of the depolarizing field and a V-shaped electrostatic potential well could be formed (Figure 10b). [182] This V-shaped potential was superimposed on the electronic potential determined by the bands off-set of the ferroelectric/dielectric layer. For BTO/STO composite barrier without a band offset,[36] an asymmetric triangular well could be formed at the ferroelectric/dielectric interface and a resonant-type peak at a certain incident energy was observed (Figure 10c).[195]

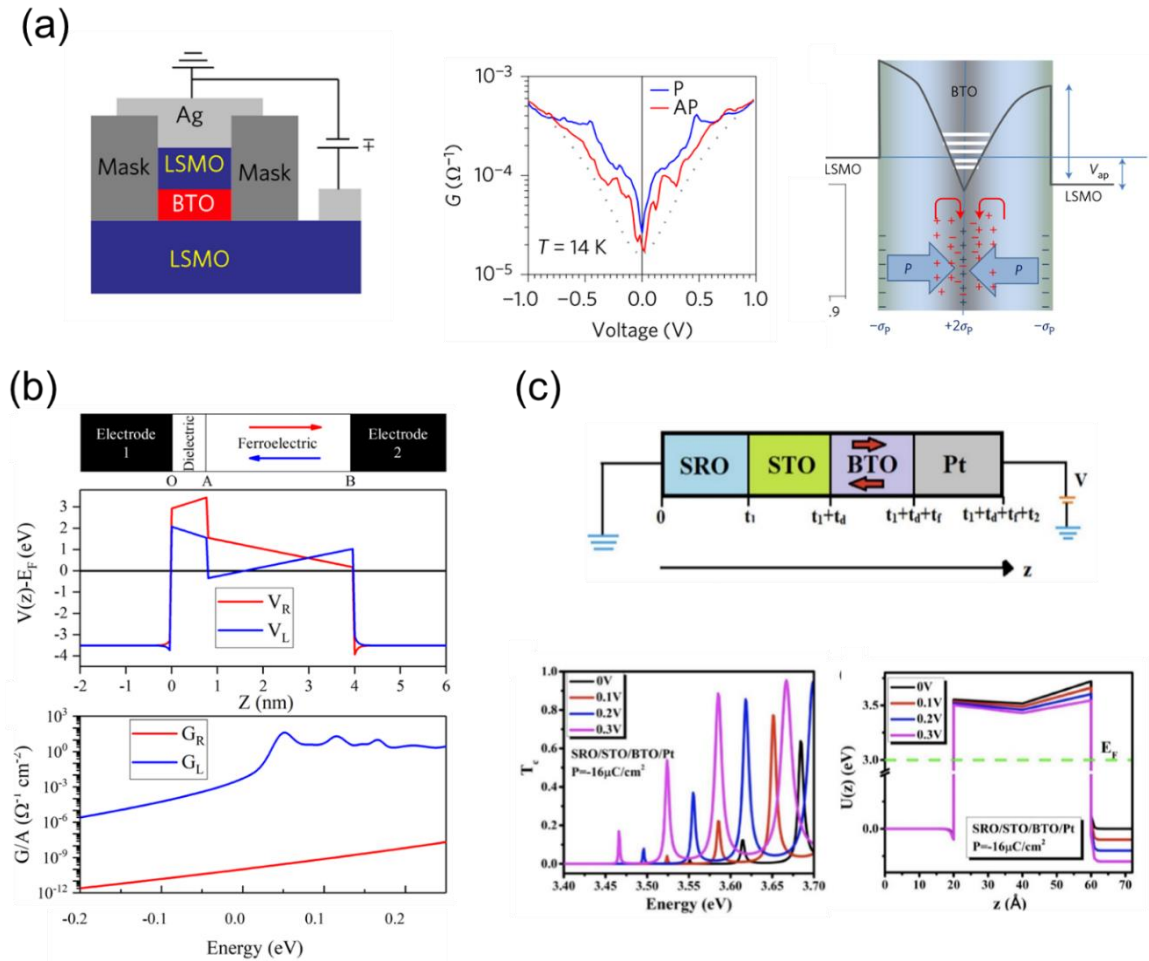

**Figure 10. a) Sketch of the structure of LSMO/BTO/LSMO MFTJs (left panel), and the differential conductance obtained as the numerical derivative of the current versus voltage for parallel (blue curve) and antiparallel (red curve) magnetic states at 14 K (middle panel), and sketch of voltage-dependent band bending and confined electronic states at the head-to-head domain wall (right panel). Reproduced with permission.[192] Copyright 2017, Springer Nature. b) Potential profiles (middle panel) and conductance per unit area (bottom panel) of an FTJ with a composite dielectric layer of a small dielectric constant as shown on the top panel. Reproduced with permission.[182] Copyright 2022, the Author(s), AIP publishing. c) Schematic presentation of an FTJ with a composite barrier (20 Å)SRO/(20 Å)STO/(20 Å)BTO/(12 Å)Pt (top panel), with shifting of quasi-resonant peaks towards lower energy values as the bias increases (bottom left panel), and the potential energy profile of the ON state demonstrates lowering of the vertex of the triangular well with an increase in bias, thus reducing the asymmetry of the triangular well (bottom right panel). Reproduced with permission.[195] Copyright 2023, Taylor & Francis Group, LLC.**

## 2.6 Other key factors in the performance of FeRTDs

In addition to the above-mentioned selection of barrier and well materials, the performance of FeRTDs is also strongly dependent on the heterostructure configuration, including the layers of the QW, thicknesses of barrier and well layers, etc.

**a. FeRTD configuration**

Experimental studies have demonstrated that in a conventional three-layer heterostructure, the NDR response resembled that of a system exhibiting resonant tunneling.[188] This behavior was observed when utilizing a ferroelectric material as the barriers and a potential well with a distinctly different bandgap. In contrast, as mentioned previously, using a ferroelectric material as a well, rather than as a barrier, resulted in a different system.[180] Alternatively, including a metallic material as the well has shown to have enhanced the modulation of resonant tunneling properties.[188]

Beyond the double barrier structure, resonant tunneling in multiple GaAs-based QW heterostructures has been proposed.[196,197] Meanwhile, theoretical insights into further modulation of resonant tunneling effects have been explored through the number of QW barriers.[198,199] It was found that increasing the number of QW layers enhanced TER, driven by the competition between the suppression of direct, and the reinforcement of resonant tunneling. Notably, both works highlighted that the fabrication of multi-QW FeRTDs, particularly the triple-QW heterostructure was superior to the conventional single-QW FeRTD (Figure 11a).

**b. Thicknesses of barrier and well**

We had previously shown that resonant tunneling is present across various thicknesses of the SRO well, ranging from 1.5 nm, 3 nm and 4.5 nm, and a large OFF/ON ratio of $2\times10^4$ is obtained for 4.5 nm SRO (Figure 11b).[188] Chang *et al*.[184,198] predicted that contributions from increasing barrier and well layer thicknesses would lead to weaker direct tunneling on $HfO_2$-based resonant tunneling diodes. For a thicker ferroelectric barrier, quantum confinement was increased, resulting in a sharper resonance and thus a stronger resonant tunneling effect. On the contrary, if the well thickness increased, the confinement was reduced due to a wider well. However, the increased number of confined states and the reduced energy difference between the states led to an overall enhancement of resonant tunneling to total current injection.

**c. Electrode selection**

The selection of electrode material significantly affects the asymmetry of FTJs and thus their TER.

Typically, metals are often used as the top electrodes. For example, Pt as a popular candidate have been used in FTJs that exhibit a giant TER.[24,137] More specific work on the impact of top electrodes was conducted by Boyn *et al*. who found that FTJs with large work function metal top electrodes would produce higher TER ratios.[200] However, this came at the cost of unreliable switching, suggesting a compromise between the two competing effects to maximize performance.

For bottom electrodes, perovskite metal oxides are usually employed to maintain the epitaxial growth of adjacent ferroelectric layers. In FeRTDs with an NSTO substrate used as the bottom electrode, at the n-type semiconductor/ferroelectric interface, either accumulation or depletion of electrons could occur depending on the direction of polarization. Thus, in response to polarization reversal, significant changes to the barrier width and height could occur, contributing to the large TER.[136]

For more detailed information on the influence of electrodes on TER, we direct the reader to our previous review[18] wherein we discuss the relevant electrode engineering methods, including epitaxial strain, electrode thickness, polarization modulation, etc.

**2.7 Other oxide based RTDs**

In addition to ferroelectrics, other oxide based RTDs have been reported. For example, STO/GdTiO$_3$/STO heterostructures showed resonant tunneling features and are used to probe sub-band spacings in 2DEGs at the oxide interfaces.[201] The presence of a 2DEG at the interface between insulating oxides have been predicted as a possible means to facilitate resonant tunneling in oxide heterostructures.[202] Choi *et al*. studied the transport properties across a QW superlattice engineered by interposing extremely thin layers of LaTiO$_3$ within the STO barrier layers, which formed a 2DEG quantum well at the interface (Figure 11c-i, ii, iii).[189] A distinct manifestation of resonant tunneling accompanied by NDR was identified (Figure 11c-iv). This resonant tunneling feature significantly enhanced the ON/OFF ratio through the bipolar resistance switching.

Recently, a new method was proposed for the tuning of the ratio of the Coulomb interaction to the band width, and hence the metal-to-insulator transition (MIT), using the resonant tunneling effect in double-QW heterostructures of strongly correlated oxides. Such double-QW heterostructure consisted of a barrier layer that was sandwiched between two QW layers. For example, Yukawa *et al*. reported double-QW heterostructures that used layers of the strongly correlated conductive oxide SrVO$_3$ and a barrier layer of the band insulator STO. Here the top QW was a marginal Mott-insulating SrVO$_3$ layer while the bottom QW was a metallic layer of SrVO$_3$ (Figure 11d).[203] With this double-QW heterostructure, the authors suggested that the MIT control by resonant tunneling was more advantageous than field-effect transistor control in correlated oxide systems.[204,205]

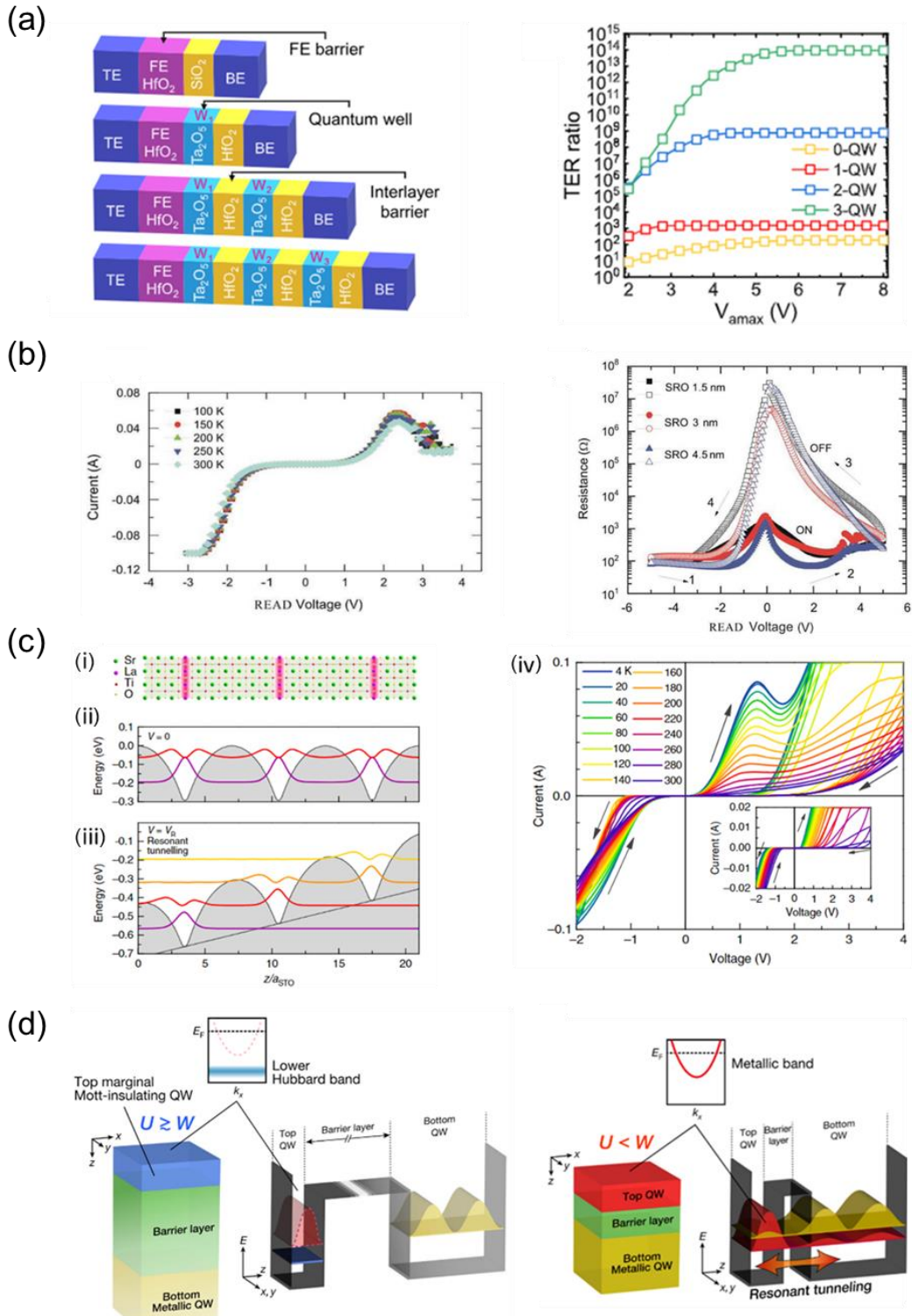

**Figure 11.** a) Schematic diagram of HfO$_2$-based multi-QW heterostructures (left panel) and their TER ratios showing significantly enhanced values for a triple-QW heterostructure (right panel). Reproduced with permission.[198] Copyright 2024, Institute of Electrical and Electronics Engineers. b) Resonant tunneling and NDR across a range of temperatures, including at room temperature (left panel), and OFF/ON resistance ratios for a BTO/STO/BTO FeRTD with differing thicknesses of the SRO well (right panel). Reproduced with permission.[188] Copyright 2022, Wiley-VCH Verlag GmbH and Co. c) Schematic diagram of a LaTiO$_3$/STO quantum oxide superlattice (i), calculated potential

wells (thin black lines) and probability functions of finding electrons (thick lines) at V = 0 (ii) and V = $V_R$ (resonant bias) (iii), and I-V characteristics demonstrating observation of resonant tunneling and NDR (iv). Reproduced under the terms of the Creative Commons CC BY license.[189] Copyright 2015, The Author(s), published by Springer Nature. d) Schematic diagram of a $SrVO_3$/STO/$SrVO_3$ double-QW heterostructure. The top $SrVO_3$ is a marginal Mott-insulating QW when its Coulomb interaction (*U*) is greater than the bandwidth (*W*), leading to localized quantized electron states (left panel). Upon resonant tunneling, quantum states between the top $SrVO_3$ and the bottom metallic $SrVO_3$ QW layers are hybridized. This results in bonding (red curve) and antibonding (yellow curve) states that allow electrons in the top QW to move to the bottom QW (right panel). Reproduced under the terms of the Creative Commons CC BY license.[203] Copyright 2021, The Author(s), published by Springer Nature.

## 3 Challenges and opportunities for hybrid junctions

Compared to conventional FTJs with a metal/ferroelectric/metal structure, ferroelectric hybrid tunneling junctions and FeRTDs provide more apparent advantages such as versatile design, enhanced TER (OFF/ON ratio), tunable functionalities, multi-resistance states, high-speed switching and so on. This opens possibilities for advanced computing architectures, such as neuromorphic computing and quantum information processing, etc. which are not available in conventional FTJs. Nevertheless, there are still some issues to be addressed for their practical applications.

Like conventional FTJs, ferroelectric hybrid tunneling junctions face challenges such as stability and scalability when used as non-volatile memories. Till now, most FTJs and MFTJs have been based on perovskite materials (including films and substrates) to facilitate the epitaxial growth of film and electrode layers.[32,128,206,207] Due to the requirements of low-energy consumption and stability of the polarization state, the tunneling current needs to be read at a small voltage, typically below 0.5 V.[124,208] Unfortunately, a low ON-state current density or large RA product has often been observed (especially for MFTJs), which was detrimental to fast data reading.[34,90,101] This problem was more pronounced for FTJs/MFTJs with a composite barrier. To achieve higher tunneling current, decreasing the ferroelectric thickness often poses the challenge of maintaining ferroelectric stability. Moreover, reducing the thickness of the ferroelectric or dielectric in FTJs with a ferroelectric/dielectric barrier easily leads to a lower TER, owing to the changes of the profile of potential energy. Exploring an optimal thickness distribution of the composite barrier could balance the current level and TER. For FTJs with asymmetric electrodes, it was probable to achieve both a relatively large ON state current and a high TER by fixing ferroelectric thickness and decreasing dielectric thickness. This approach ensured stable ferroelectricity and enhanced electron tunneling probability. The asymmetry of the potential profile might not change monotonously with the dielectric thickness, unlike the case of symmetric electrodes.

Besides barrier width, the band offset of the ferroelectric/dielectric layer also greatly influences the asymmetry of the potential energy profile. A large band offset causes significant asymmetry and consequently a large TER. A giant TER was achieved theoretically in FTJs with a MgO/BTO composite barrier due to the large difference in barrier height between BTO and MgO.[36] However, the epitaxial growth of ultrathin MgO thin films remains a challenge. Additionally, the large barrier height of the dielectric layer (*e.g.,* MgO) was not advantageous to achieving a large ON state current. Band engineering of FTJs using a dielectric layer with a low barrier height provides a viable way to address the tradeoff between the ON state current and TER.

2D vdW ferroelectrics offer a promising solution to the stability issue of ferroelectricity existing in conventional ferroelectrics, such as the critical size and mandatory charge screening. In addition, the tunable energy band structures of 2D vdW ferroelectrics[209] may solve the problem of a large RA product in perovskite-based MFTJs if an ultrathin 2D vdW ferroelectric with a relatively narrower energy gap were to be used as the tunnel barrier. Su *et al.* proposed vdW MFTJs consisting of 2D ferromagnetic and 2D ferroelectric materials ($Fe_mGeTe_2/\alpha$-$In_2Se_3/Fe_nGeTe_2$; m, n =3, 4, 5; m ≠ n).[90] In these MFTJs, a remarkably low RA product less than 1 $\Omega \cdot \mu m^2$ was found, which could be attributed to the small bandgap of $\alpha$-$In_2Se_3$.

In superconducting junctions where YBCO acts as an electrode, electrochemistry could in some cases account for the electric driven resistance switching due to the high reduction potential of YBCO. For example, in YBCO/Al junctions without a ferroelectric barrier, the oxidation state could be manipulated through an external electric field, resulting in reversible resistive switching.[166] Notably, the YBCO/Al junction exhibited non-volatile memristive behavior where the conductance could be controlled by the history of applied gate pulses, achieving an ON/OFF ratio as high as 10,000%.[166] In these systems, the effect was explained by the reversible redox reaction with the motion of oxygen ions in the CuO chains. Holes could also move when stimulated by electric field or illumination. Villegas *et al.* demonstrated an additional optical response could be observed in the YBCO/ITO junction in either ON or OFF states after electrical switching (*i.e.,* OFF and ON states) upon illumination.[165] This phenomenon was attributed to the movement of oxygen or holes. Specifically, at the ON state, relaxation towards its OFF ground states was notably accelerated, likely due to ion diffusion activated by phonon excitation through light absorption. Conversely, in the OFF state, there was a large open circuit voltage due to the photovoltaic effect, leading to a hole accumulation at the YBCO interface. This hole accumulation prompted the oxidation of Cu ($Cu^{+2} + h^+ \rightarrow Cu^{+3}$), resulting in a thinning of the oxygen depleted YBCO layer and thereby enhancing electrical conductance towards the ON state.[165]

Although several investigations of the ferroelectric field effect on superconductivity have been conducted, research on ferroelectric/superconducting tunnel junctions still remains limited.[151,155,210] In MoSi/BFO/YBCO tunnel junctions, the redox reaction might be largely overcome if the MoSi electrode was replaced with Au, *i.e.,* preventing the oxygen migration between electrodes. The ferroelectric field effect also might play the key role in resistance switching in ferroelectric/superconducting tunnel junctions by employing a $SrCuO_2$ electrode. Unlike YBCO, all the atoms in $SrCuO_2$ are strongly bonded, therefore the resistance change with polarization reversal can only arise from changes in the free charge carrier density in the material.[211] In addition, the phenomenon described in YBCO/ITO junctions could introduce an additional degree of freedom in tunning the transport behavior by establishing an ultrathin ferroelectric barrier, to achieve optically controlled electroresistance and electrically controlled photovoltage.[212] Recently, theoretical and experimental investigations have revealed electrically tunable superconductivity in 2D vdW materials, such as monolayer $WTe_2$ and bilayer $T_d$-$MoTe_2$.[213–215] 2D vdW FTJs with a superconducting electrode or barrier could provide the possibility to explore the TER effect and the interplay between superconductivity and ferroelectricity in 2D materials. In these 2D SC/FE hybrid tunnel junctions, with rational design, a large TER might be expected due to the parallel modulation of the barrier height and superconductivity by ferroelectric polarization.

Attention in ferroelectrics has primarily focused on ferroelectric memories, with FTJs emerging as the leading example of ferroelectric memristors.[53,54,216,217] This field is fueled by the urgent demand to address computing limitations such as the von Neumann bottleneck[218] and the approaching limit to Moore's law.[219] In response to the current state, there has been a paradigm shift in research towards applications of ferroelectrics in machine learning and neuromorphic (brain-inspired) computing.[29,220–222] This growing interest also extends to FeRTDs, particularly regarding their practical applications. The non-volatility in ferroelectric memories and enhanced TER values associated with resonant tunneling, as opposed to direct tunneling, should not be overlooked. Advances in equipment and techniques for ferroelectric thin film fabrication, along with the discovery of ferroelectricity in ultrathin films [128,223] further facilitate the progress in this area.

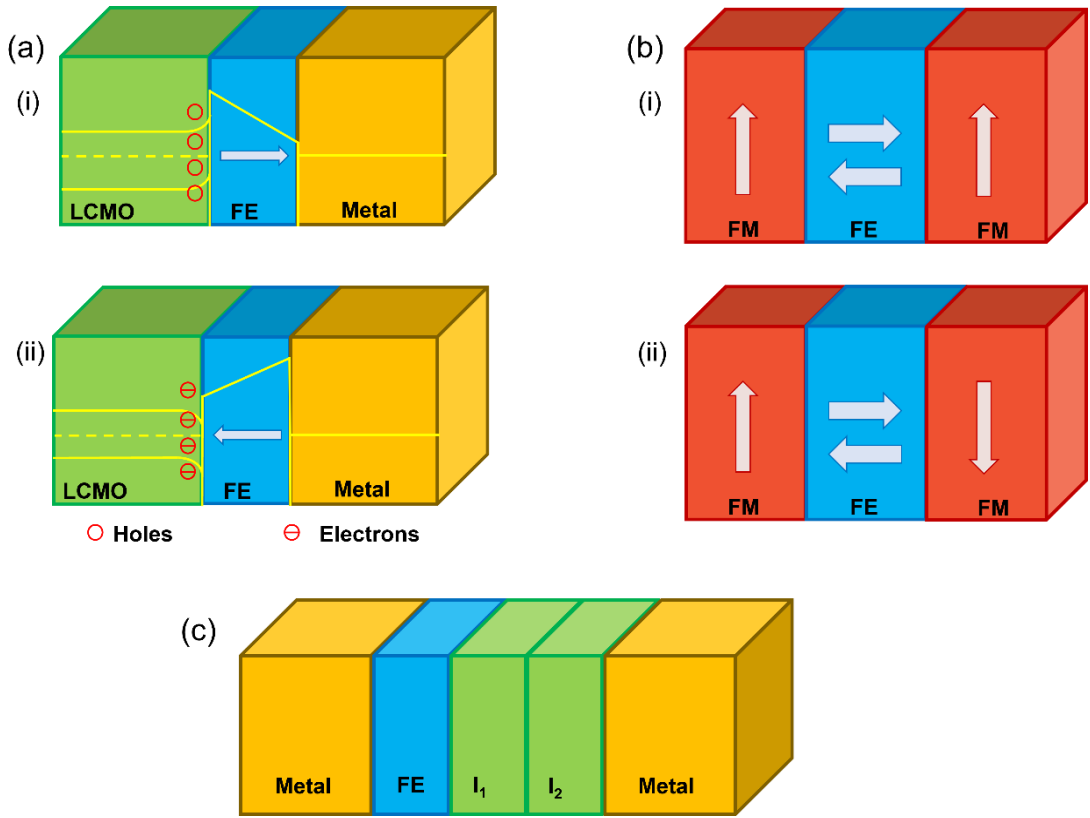

**Figure 12.** Schematic diagrams illustrating respective perspectives of tunneling heterostructures. a) At the La$_{1-x}$Cr$_x$MnO$_3$ (LCMO)/FE interface, dependent on the direction of the ferroelectric, the MIT induces a change of LCMO to metallic when there is hole accumulation (i) to insulating when there is hole depletion (ii). b) In a prototypical MFTJ, for either polarization direction the spins could be either parallel (i) or anti-parallel (ii), resulting in four possible resistance states. c) The proposed representation of a FE/I$_1$/I$_2$ QW heterostructure.[226]

The engineering of FeRTDs is at a nascent stage to realize above-mentioned applications, much more work can be done in the future with an exciting outlook towards improved performance. With its structure similar to FTJs and hybrid junctions, inspiration can be taken to emulate novel approaches. The insertion of a correlated electron oxide at the interface between a ferroelectric has been shown to increase asymmetry within an FTJ.[31] This occurs due to the MIT induced by ferroelectric polarization switching (Figure 12a). Perhaps, using one such material, La$_{1-x}$A$_x$MnO$_3$ (A = Sr, Ca) as the QW layer could increase TER effects in a FeRTD. This notion has been shown to enhance the TER effect in an artificial MFTJ, where a correlated electron oxide is placed in between a ferromagnetic electrode and ferroelectric barrier.[32] Relative to FTJs, MFTJs introduces an additional set of two resistance states (Figure 12b), where the TMR depends on the orientation of the ferromagnetic spins (parallel or anti-parallel).[224] This co-existence of TER and TMR results in four resistance states,[93] increasing the prospect for greater functionality in storage applications. By utilizing the interfacial magnetoelectric coupling between a ferromagnetic and ferroelectric interface, resistance states can be modulated by electric control of spins,[81,101,106,225] and potentially the magnetic control of polarization. Translation of this concept to a FeRTD with a ferromagnetic electrode may lead to a multiferroic resonant tunneling diode with four different resistance states. Another possible proposal is a FE/I$_1$/I$_2$ QW heterostructure .[226] Here, I$_1$ is a low-barrier dielectric that acts as the QW when sandwiched between a ferroelectric and an insulator I$_2$ (Figure 12c) with a different barrier height and permittivity to the ferroelectric. Regardless of what route follows for FeRTDs, there are a plethora of options available to be explored.

## 4  Summary

In conclusion, hybrid FTJs and FeRTDs hold immense potential for low-energy, high-speed electronic devices, addressing the growing demands in computing and information processing. Compared with conventional FTJs, hybrid FTJs offer additional possibilities to modulate the tunneling phenomena and performance, potentially expanding their applications. In hybrid FTJs, the TER can be enhanced through additional control over the selection of partner materials, *i.e.,* dielectric, 2D materials, and/or ferromagnetic and superconducting materials. FeRTDs and associated NDR behaviors can be precisely controlled through ferroelectric polarization, making them beneficial for advanced electronics with tunable and highly responsive functionalities. We hope this review will further inspire the development of novel tunneling devices and stimulate ongoing research on hybrid FTJs and FeRTDs for mainstream electronic applications.

**Acknowledgements**